\documentstyle[editedvolume]{crckapb}

\input psfig.tex

\newcommand{\be}{\begin{equation}}
\newcommand{\ee}{\end{equation}}
\def\lsim{\lower.5ex\hbox{$\; \buildrel < \over \sim \;$}}
\def\gsim{\lower.5ex\hbox{$\; \buildrel > \over \sim \;$}}

\begin{opening}
\title{Nucleosynthesis Around Black Holes}

\author{Banibrata Mukhopadhyay}
\institute{Theoretical Astrophysics Group,\\
	S. N. Bose National Centre For Basic Sciences
	JD Block, Salt Lake, Sector-III, Calcutta-700091,
	India\\
	e-mail: bm@boson.bose.res.in}

\end{opening}

\runningtitle{Nucleosynthesis Around Black Holes }

\begin{document}
\begin{abstract}
{\small 
Study of nucleosynthesis in accretion disks around
black holes was initiated by Chakrabarti et al. (1987).
In the present work we do the similar analysis
using the state-of-the-art disk model, namely,
Advective Accretion Disks. During the infall,
matter temperature and density are generally increased
which are first computed. These quantities are used to
obtain local changes in composition, amount of nuclear
energy released or absorbed, etc. under various inflow
conditions. In the cases where the magnetic viscosity is             
dominant neutron torus may be formed. We also talk about
the fate of $Li^7$ and $D$ during the accretion.
The outflowing winds from the disk could carry the new
isotopes produced by nucleosynthesis and contaminate
the surroundings. From the degree of contamination,
one could pinpoint the inflow parameters.}
\end{abstract}
\smallskip
{\small

}

\section{Introduction}

There are many observational evidences where the incoming 
matter has the potential to become
as hot as its virial temperature $T_{virial} \sim 10^{13}$ $K$ 
[1]. Through various cooling effects,
this incoming matter is usually cooled down to produce hard and soft states [2].
In the accretion disk, matter in the
sub-Keplerian region generally remains hotter than Keplerian disks. 
The matter is so hot that after big-bang nucleosynthesis this is the most
fevourable temperature to produce significant nuclear reactions.
The energy generation due to nucleosynthesis could be high
enough to destabilize the flow and the modified composition
may come out through winds to affect the metallicity of the galaxy
[3-7]. Previous works on nucleosynthesis in disk
was done for cooler thick accretion disks.
Since the sub-Keplerian region is much hotter
than of Keplerian region and also than the central temperature ($\sim 10^7$K)
of stars, presently we are interested to study 
nucleosynthesis in hot sub-Keplerian region of accretion disks.

\section{Basic Equations and Physical Systems} 

In 1981 Paczy\'nski \& Bisnovatyi-Kogan [8] initiated the study of 
viscous transonic flow although the global solutions of advective accretion disks 
were obtained much later [9] which we use here. In the advective disks, 
matter must have radial motion which is transonic. The supersonic 
flow must be sub-Keplerian and therefore must deviate from a
Keplerian disk away from the black hole. The basic equations 
which matter obeys while falling towards the black hole 
from the boundary between Keplerian and sub-Keplerian
region are given below (for details, see, [9]):

\noindent (a) The radial momentum equation:

$$
\vartheta \frac{d\vartheta}{dx} +\frac{1}{\rho}\frac{dP}{dx}
+\frac {\lambda_{Kep}^2-\lambda^2}{x^3}=0,
\eqno{(1a)}
$$

\noindent (b) The continuity equation:

$$
\frac{d}{dx} (\Sigma x \vartheta) =0 ,
\eqno{(1b)}
$$

\noindent (c) The azimuthal momentum equation: 

$$
\vartheta\frac{d \lambda(x)}{dx} -\frac {1}{\Sigma x}\frac{d}{dx}
(x^2 W_{x\phi}) =0 ,
\eqno{(1c)}
$$

\noindent (d) The entropy equation:

$$
\frac{2na{\rho}{\vartheta}h(x)}{\gamma}
\frac{da}{dx}-\frac{a^2{\vartheta}h(x)}
{\gamma}\frac{d{\rho}}{dx}=fQ^+
\eqno{(1d)}
$$
where the equation of state is chosen as $a^2=\frac{\gamma P}{\rho}$.
Here, $\lambda$ is the specific angular momentum of the infalling matter, 
$\lambda_{Kep}$ is that in the Keplerian region is defined as
$\lambda_{Kep}^2 = \frac{x^3}{2(x-1)^2}$ [10], $\Sigma$ is vertically integrated 
density, $W_{x\phi}$ is the stress tensor, $a$ is the sound speed and
$h(x)$ is the half thickness of the disk ($\sim ax^{1/2}(x-1)$),
$n = \frac{1}{\gamma-1}$ is the polytropic index, $f$ is the cooling 
factor which is kept constant throughout our study, $Q^+$ is the
heat generation due to the viscous effect of the disk. For the time
being we are neglecting the magnetic heating term. 

During infall different nuclear reactions take place and nuclear
energy is released. Here, our study is exploratory so in the heating term $Q^+$, we
do not include the heating due to nuclear reactions. (Work including nuclear energy release 
term is in [6].) Another parameter $\beta$ is defined as ratio of gas pressure to
total pressure, which is assumed to be a constant value throughout a
particular case. Actually, the factor $\beta$ is used to take into
account the cooling due to Comptonization.
To compute the temperature of the Comptonized flow in the advective region
which may or may not have shocks, we follow Chakrabarti \& Titarchuk [2] and 
Chakrabarti's [11] works and method. The temperature is computed from.
$$
T=\frac{a^2{\mu}{m_p}{\beta}}{{\gamma}k}.
\eqno{(2)}
$$
It is seen that due to hotter nature of the advective disk 
especially when accretion rate is low, 
Compton cooling is negligible, the major precess of hydrogen burning
is the rapid proton capture process, which operates at
$T \gsim 0.5 \times 10^9$K which is much higher than the operating temperature
of PP chain (operates at $T \sim 0.01-0.2 \times 10^9$K) and CNO
cycle (operates at $T \sim 0.02-0.5 \times 10^9$K) which take place
in the case of stellar nucleosynthesis where temperature is much lower.
Also in stellar case, in different radii same sets of reaction take
place but in the case of disk, in different radii different reactions
(or different sets of reaction) can take place simultaneously. 
These are the basic differences between the nucleosynthesis 
in stars and disks.   

For simplicity, we take the solar abundance as the initial abundance
of the disk and our computation starts where matter leaves a Keplerian disk.
According to [2] and [11], the black hole remains in hard states
when viscosity and accretion rate are smaller. In this case,
$x_K$ (at radius $x_K$ matter deviates from Keplerian to sub-Keplerian
region) is large. In this parameter range the protons remain hot 
($T_p \sim 1 - 10 \times 10^9$K). The corresponding factor
$f (= 1 - Q^+/Q^-)$ is not low enough to cool down the disk, 
(in [1], it is indicated that $\dot{m}/\alpha^2$ is a good indication 
of the cooling efficiency of the hot flow), where $Q^+$ and $Q^-$ 
are the heat gain and heat loss due to viscosity of the disk.

We have studied a large region of parameter space with
$0.0001 \lsim \alpha \lsim 1$, $0.001\lsim {\dot m} 
\lsim 100$, $0.01 \lsim \beta \lsim 1$, $4/3 \lsim \gamma 
\lsim 5/3$. We study a case with a standing shock as well.
In selecting the reaction network we kept in mind the fact that hotter
flows may produce heavier elements through triple-$\alpha$ and rapid
proton and $\alpha$ capture processes. Furthermore due to photo-dissociation
significant neutrons may be produced and there is a possibility of
production of neutron rich isotopes. Thus, we consider sufficient number
of isotopes on either side of the stability line. The network thus
contains protons, neutrons, till $^{72}Ge$ -- altogether 255 nuclear
species. The standard reaction rates were taken [6].

\section{Results}

Here we present a typical case containing a shock wave
in the advective region [6]. We express the length in the unit of
one Schwarzschild radius which is $\frac{2GM}{c^2}$
where $M$ is the mass of the black hole, velocity is expressed
in the unit of velocity of light $c$ and the unit of time
is $\frac{2GM}{c^3}$. We use the  mass of the black hole
$M/M_\odot=10$ ($M_{\odot}=$ solar mass), 
$\Pi$-stress viscosity parameter $\alpha_\Pi=0.05$,
the location of the inner sonic point $x_{in}=2.8695$, the 
value of the specific angular
momentum at the inner edge of the black hole $\lambda_{in}=1.6$, 
the polytropic index $\gamma=4/3$ 
as free parameters. The net accretion rate ${\dot m}=1$
in the unit of Eddington rate, cooling factor due
to Comptonization $\beta = 0.03$, $x_K=481$. 
The proton temperature (in the unit of $10^9$), velocity distribution 
(in the units of $10^{10}$ cm sec$^{-1}$), density distribution
(in the unit of $2 \times 10^{-8}$ gm cm$^{-3}$) 
are shown in Fig. 1(a).
\begin{figure}
\vbox{
\vskip -0.5cm
\hskip 0.0cm
\centerline{
\psfig{figure=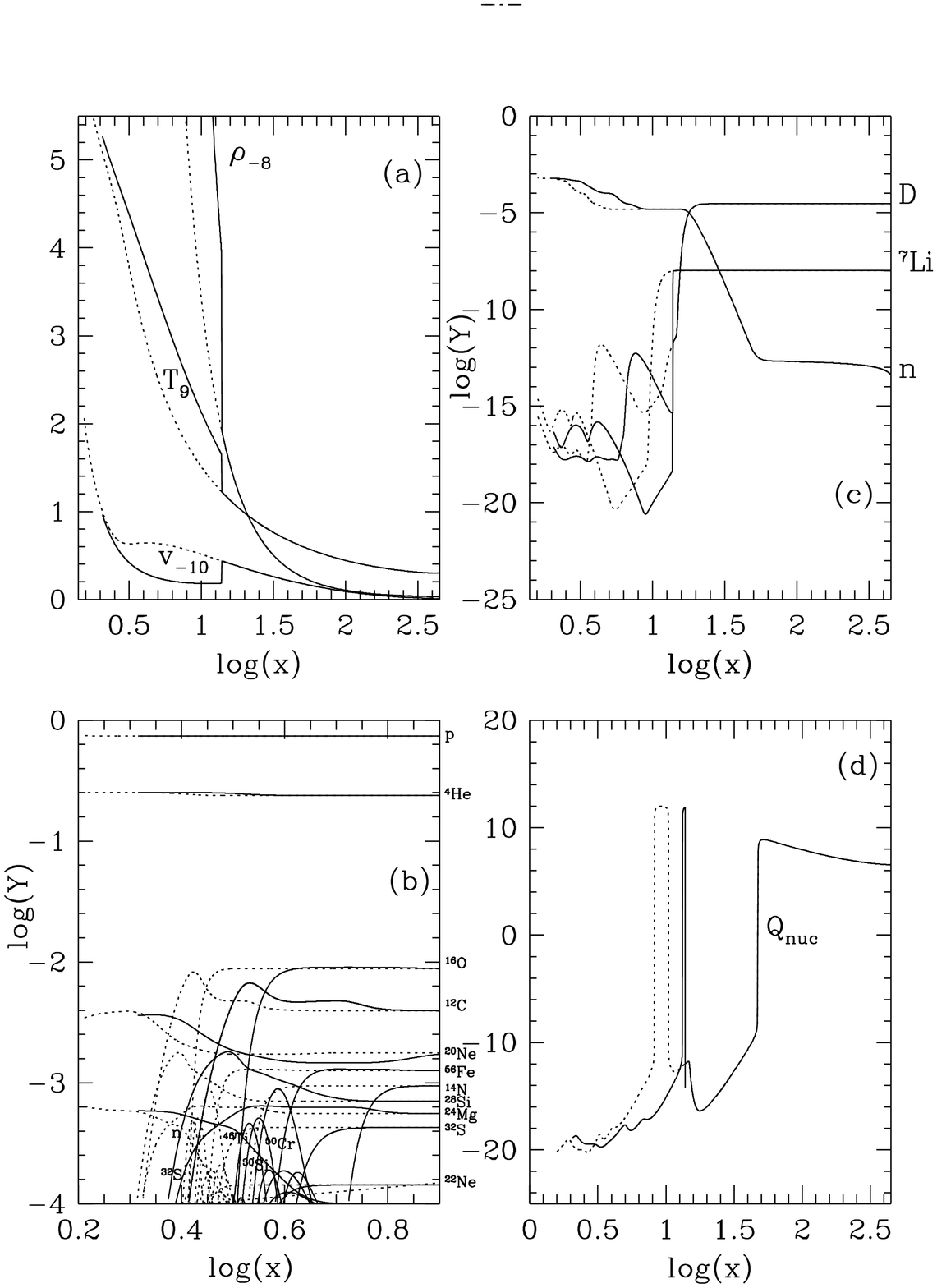,height=13truecm,width=13truecm}}}
\vspace{-0.0cm}
\noindent {\small {\bf Figure 1} : Variation of (a) proton temperature ($T_9$), radial velocity $v_{10}$ and
density distribution $\rho_{-8}$
(b)  matter abundance $Y$ in logarithmic scale (c) neutron, deuterium and lithium abundance  
$Y$ in logarithmic scale and (d) nuclear energy release
and absorption as a functions of logarithmic radial distance $x$.
See text for parameters. Solutions in the stable branch with shocks are solid curves 
and those without the shock are dotted in (a-d). 
At the shock, temperature and density rise and velocity lower significantly and cause
a significant change in abundance even farther out. Shock induced winds may
cause substantial contamination of the galactic composition when parameters are
chosen from these regions [6]. }  
\end{figure}

In Fig. 1b, we show composition changes close to the black hole
both for the shock-free branch (dotted curves) and the shocked branch
of the solution (solid curves). Only prominent elements are plotted.
The difference between the shocked and the shock-free cases is that, in the shock
case the similar burning takes place farther away from the black hole
because of much higher temperature in the post-shock region.
A significant amount of the neutron
(with a final abundance of $Y_n \sim 10^{-3}$) is produced due
to photo-dissociation process. Note that closer to the black hole,
$^{12}\!C$, $^{16}\!O$, $^{24}\!Mg$ and $^{28}\!Si$ are all destroyed
completely. Among the new species which are formed
closer to the black hole are $^{30}\!Si$, $^{46}\!Ti$, $^{50}\!Cr$. Note that
the final abundance of $^{20}\!Ne$ is significantly higher than the initial value. Thus a significant
metallicity could be supplied by winds from the centrifugal barrier.  
In Fig. 1c we show the change of abundance of neutron ($n$), deuterium ($D$) and
lithium ($^{7}\!Li$). It is noted that near black hole a significant
amount of neutron is formed although initially neutron abundance was
almost zero. Also $D$ and $^{7}\!Li$ are totally burnt out near
black hole which is against the major claim of Yi \& Narayan [13] which 
found significant lithium in the disk. It is true that due to
spallation reaction, i.e.,
$$
^{4}\!He+^{4}\!He \rightarrow ^{7}\!Li + p
$$
$^{7}\!Li$ may be formed in the disk but due to photo-dissociation in 
high temperature all $^{4}\!He$ are burnt out before forming $^{7}\!Li$
i.e. the formation rate of $^{4}\!He$ from $D$ is much slower than
the burning rate of it. Yi \& Narayan [13] do not include the possibility
of photo-dissociation in the hot disk. 
 
In Fig. 1d, we show nuclear energy release/absorption for the flow in
in units of erg sec$^{-1}$ gm$^{-1}$. Solid curve represents the 
nuclear energy release/absorption for the shocked flow 
and the dotted curve is that for unstable shock-free flow. 
As matter leaves the Keplerian region, the rapid proton capture
such as, $p+ ^{18}\!O \rightarrow  ^{15}\!N + ^{4}\!He$ etc.,
burn hydrogen and releases energy to the disk.
At around $x=50$, $D \rightarrow n + p $ dissociates $D$
and the endothermic reaction causes the nuclear energy release to become 
`negative', i.e., a huge amount of energy is absorbed from the disk. 
At around $x=15$ the energy release is again dominated by the  original 
processes because no deuterium is left to burn.
Due to excessive temperature, immediately $^3\!He$ breaks down into deuterium
and through dissociation of $D$ again a huge amount of energy is absorbed
from the disk.  It is noted that energy absorption due to photo-dissociation
as well as the magnitude of the energy release due to proton capture process and
that due to viscous dissipation ($Q^+$) are {\it very} similar
(save the region where endothermic reactions dominate). 
This suggests that even with nuclear reactions, 
at least some part of the advective disk may be perfectly stable.

\begin{figure}
\vbox{
\vskip -2.0cm
\hskip 22.0cm
\centerline{
\psfig{figure=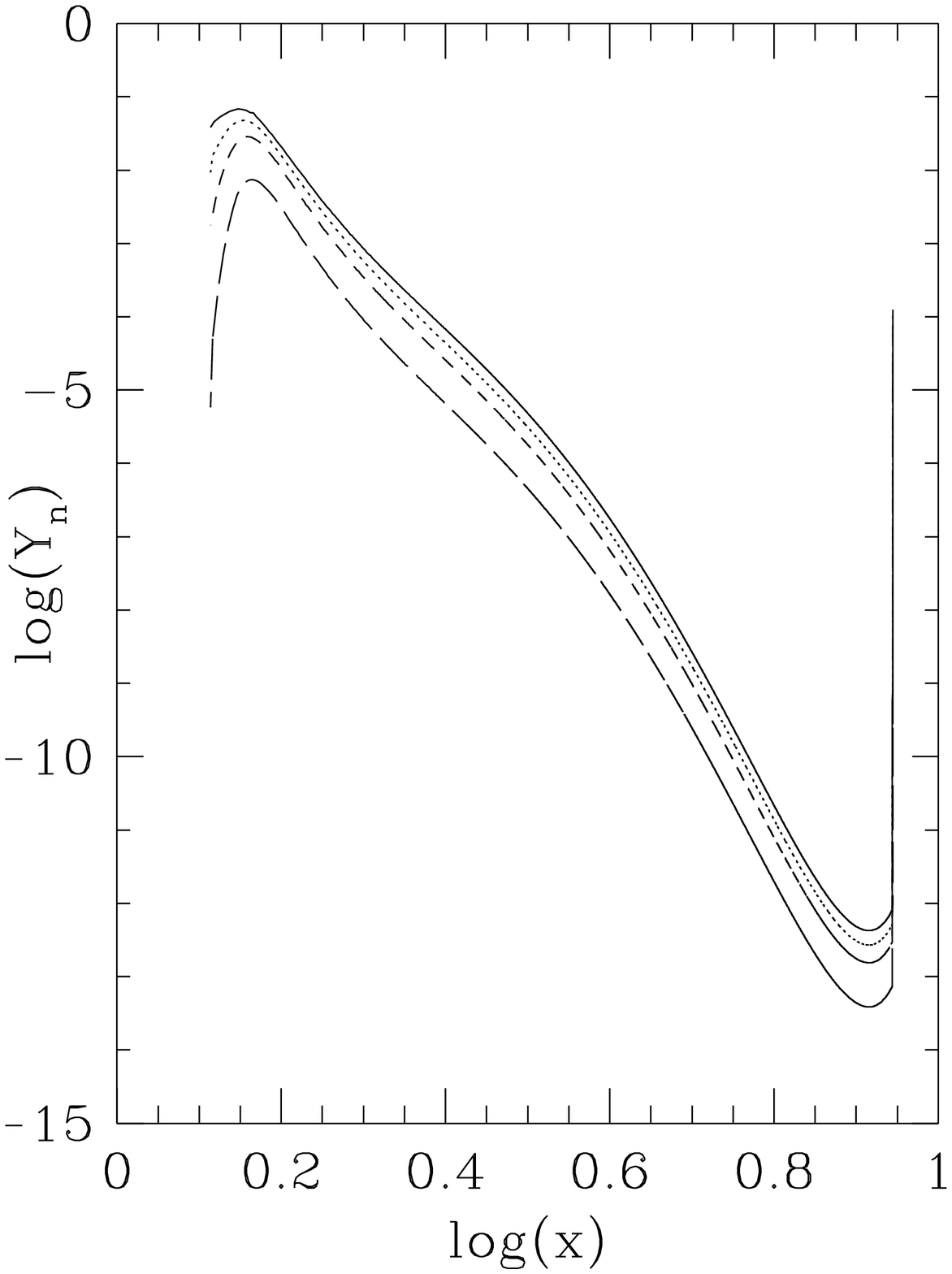,height=10truecm,width=10truecm}}}
\vspace{-0.0cm}
\noindent {\small {\bf Figure 2 :}
The convergence of the neutron abundance through successive iterations in a
very hot advective disk. From bottom to top curves $1$st, $4$th, $7$th and 
$11$th iteration results are shown. A neutron torus with a significant abundance is
formed in this case [15]. }
\end{figure}

We now present another interesting case where lower accretion rate
(${\dot m}=0.01$) but higher viscosity ($0.2$) were used and 
the efficiency of cooling
is not $100\%$ ($f=0.1$). That means that the temperature of the flow is high
($\beta = 0.1$, maximum temperature $T_9^{max}=11$). In this case $x_K=8.8$, 
if the high viscosity is due to stochastic
magnetic field, protons would be drifted towards the black hole
due to magnetic viscosity, but the neutrons will not be
drifted [13] till they decay. This principle has been used
to do the simulation in this case. The modified composition in one
sweep is allowed to interact with freshly
accreting matter with the understanding that the accumulated neutrons
do not drift radially. After few iterations or sweeps the steady
distribution of the composition may be achieved. Figure 2
shows the neutron distributions in iteration numbers
$1$, $4$, $7$ \& $11$ respectively (from bottom to top curves)
in the advective region. The formation
of a `neutron torus' is very apparent in this result and generally in all
the hot advective flows. In 1987 Hogan \& Applegate [14] showed that
formation of neutron torus is possible with high accretion
rate. But high accretion rate means high rate of photon to dump into
sub-Keplerian region and high rate of inverse Compton process 
through which matter
cool down, that is why photo-dissociation will be less prominent. Also
formation of neutron is possible through the photo-dissociation of 
deuterium in the hot disk which is physically possible 
prominently in our parameter region, where neutron torus is formed. 
Details are in Chakrabarti \& Mukhopadhyay [15].

\section{Discussions and Conclusions}

In this paper, we have explored the possibility of nuclear reactions in
advective accretion flows around black holes. 
Temperature in this region is controlled by the 
efficiencies of bremsstrahlung and Comptonization processes [2, 7]. 
For a higher Keplerian rate and higher viscosity, the inner edge of the Keplerian
component comes closer to the black hole and the advective region becomes
cooler [2, 9]. However, as the viscosity is decreased, the inner edge of the Keplerian 
component moves away and the Compton cooling becomes less efficient.

The composition changes especially in the centrifugal pressure supported denser region,
where matter is hotter and slowly moving. Since centrifugal 
pressure supported region can be treated as an effective
surface of the black hole which may generate winds and outflows in the
same way as the stellar surface, one could envisage that the winds 
produced in this region would carry away modified composition [16-18].
In very hot disks, a significant amount of 
free neutrons are produced which, while coming out through winds may 
recombine with outflowing protons at a cooler environment
to possibly form deuteriums. A few related 
questions have been asked lately: Can lithium in the universe be 
produced in black hole accretion [12,19]?
We believe that this is not possible. 
When the full network is used we find that the hotter disks where 
spallation would have been important also heliums photo-dissociate 
into  deuteriums and then to protons and neutrons before any significant
production of lithiums. Another question is: Could the metallicity
of the galaxy be explained, at least partially, by nuclear reactions?
We believe that this is quite possible. Details are in [6].

Another important thing which we find that in the case of hot inflows
formation of neutron tori is a very distinct possibility [15].
Presence of a neutron torus around a black hole
would help the formation of neutron rich species as well, a process hitherto 
attributed to the supernovae explosions only. It can also help production of 
Li on the companion star surface (see [6] and references therein).

The advective disks as we know today do not perfectly match with a Keplerian disk. 
The shear, i.e., $d\Omega/dx$ is always very small in the advective 
flow compared to that of a Keplerian disk near the outer boundary of the 
advective region. 
Thus some improvements of the disk model at the
transition region is needed. Since major reactions are closer to the
black hole, we believe that such modifications of the model would not change
our conclusions. The neutrino luminosity in a steady disk is generally very 
small compared to the photon luminosity [6], but occasionally,
it is seen to be very high. In these cases, we predict that the disk would
be unstable. Neutrino luminosity from a cool advective disk is low. 

In all the cases, even when the nuclear composition changes are not
very significant, we note that the nuclear energy release due to exothermic
reactions or absorption of energy due to endothermic reactions is
of the same order as the gravitational binding energy release.
Like the energy release due to viscous processes,
nuclear energy release strongly depends on temperatures.
This additional energy source or sink may destabilize the flow [6].

\section{Acknowledgment}

I would like to thank Prof. Sandip K. Chakrabarti for introducing me
to this subject and helpful discussion throughout the work.

\end{document}